\documentclass[]{JHEP3}%
\usepackage{graphics}      
\usepackage{epsfig,latexsym}

\def\L{\mathcal{L}}

\newcommand{\be}{\begin{equation}}
\newcommand{\ee}{\end{equation}}
\newcommand{\bea}{\begin{eqnarray}}
\newcommand{\eea}{\end{eqnarray}}
\newcommand{\lsim}{\mbox{\raisebox{-.6ex}{~$\stackrel{<}{\sim}$~}}}
\newcommand{\gsim}{\mbox{\raisebox{-.6ex}{~$\stackrel{>}{\sim}$~}}}

\newcommand{\sH}{{\scriptscriptstyle H}}

\def\done{\delta_1}
\def\dtwo{\delta_2}
\def\sH{\mathcal{H}}

\def\lap{\vec{\nabla}^2}
\def\Vint{V_{\mathrm{int}}}
\def\Uint{U_{\mathrm{int}}}
\def\O{\mathcal{O}}
\def\L{\mathcal{L}}
\def\Jint{J_{\mathrm{int}}}
\def\Jinttwo{J_{\mathrm{int,2}}}
\def\Fbox{F\left(\frac{\Box}{m_s^2}\right)}

\def\Fpws{F'\left(-\frac{{\omega^2}}{m_s^2}\right)}
\def\Gbox{\Gamma\left(\frac{\Box}{m_s^2}\right)}
\def\Gboxzero{\Gamma\left(\frac{\Box_0}{m_s^2}\right)}
\def\Gomega{\Gamma\left(-\frac{\omega^2}{m_s^2}\right)}

\def\xieff{{\xi_{\mathrm{eff}}^{(2)}}}

\title{Predictions for Nongaussianity from Nonlocal Inflation}

\author{Neil Barnaby \\ Canadian Institute for Theoretical Astrophysics,
University of Toronto, 60 St.\ George St.\, Toronto, Ontario, Canada M5S 3H8\\
Email: \email{barnaby@cita.utoronto.ca}}

\author{James M.\ Cline\\
Physics Department, McGill University, Montr\'eal, Qu\'ebec, Canada H3A
2T8\\
E-mail: \email{jcline@physics.mcgill.ca}}

\preprint{}
\abstract{
In our previous work the nonlinearity parameter $f_{NL}$, which characterizes
nongaussianity in the cosmic microwave background, was estimated for a
class of inflationary models based on nonlocal field theory.  These models
include $p$-adic inflation and generically have the remarkable property
that slow roll inflation can proceed even with an extremely steep
potential.  Previous calculations found that large nongaussianity is
possible; however, the technical complications associated with studying
perturbations in theories with infinitely many derivatives forced us
to provide only an order of magnitude estimate for $f_{NL}$.  We reconsider
the problem of computing $f_{NL}$ in nonlocal inflation models, showing
that a particular choice of field basis and 
recent progress in cosmological perturbation theory makes an exact
computation possible.  We provide the first quantitatively accurate 
computation of the bispectrum in nonlocal inflation, confirming our previous
claim that it can be observably large.  We show that the 
shape of the bispectrum in this class of models makes it 
observationally distinguishable from Dirac-Born-Infeld  inflation models.
}


\begin{document}

\section{Introduction}

Nongaussianity in the Cosmic Microwave Background (CMB) is usually
characterized by the dimensionless nonlinearity parameter, $f_{NL}$, which is
derived from the bispectrum of the curvature perturbation.  In the simplest
models of inflation $f_{NL}$ is suppressed  by slow roll parameters and hence
one has $|f_{NL}| \ll 1$  \cite{BMR}-\cite{SeeryLidsey}.  Nongaussianity can
further be characterized using the trispectrum, which is also small in the
simplest models \cite{trispectrum}.  (See also
\cite{otherNG}-\cite{post_inf_NG2} for further details and see \cite{NGreview} for a review.)  On
the other hand, if $|f_{NL}| \gsim 5$ is observed it would rule out the
minimal inflationary scenario and favour one of a relatively small number of
nonminimal constructions.   Such models include single field inflation with a small sound speed 
\cite{large_nongauss}.  An example is Dirac-Born-Infeld (DBI) inflation \cite{DBI}, whose generalization
to multiple fields has recently been  considered in \cite{multiDBI}.  Another
way to get observably large $f_{NL}$ from single field models is by having one
or more sharp features in the inflaton potential \cite{feature}.  There has
also been considerable interest constructing multi-field models which can
potentially give large nongaussianity using canonical fields and smooth
potentials \cite{multi_field}.  Successful examples include the curvaton model
\cite{curv,curvaton} and hybrid inflation \cite{hybrid} where large nongaussianity
may be generated during the preheating phase, after inflation
\cite{BC1,BC2}.  (See also \cite{preheatNG} for further studies of
nongaussianity from preheating.)  

The current limit on nongaussianity in the CMB is $-9 < f_{NL} < 111$\footnote{This is assuming the ``local'' form of nongaussianity.  For the equilateral model
one has the slightly weaker constraint $-151 < f_{NL} < 253$.} 
for the WMAP five-year data \cite{WMAP5} and future missions are
expected to be able to probe $f_{NL}$ down to order unity \cite{future}. 
Efforts to construct inflationary models with $|f_{NL}| \gg 1$ have recently
received additional motivation from the claim of a detection of $26.9 < f_{NL} <
146.7$  at 95 \% C.L. in \cite{detection} (see also \cite{smoot}, which
claims a lower but nonvanishing detection, with central value $f_{NL}\cong 50$,
at the same confidence level).  In order to assess the significance of such detections
it will be crucial to be able to measure $f_{NL}$ using independent  techniques, 
such as through its effect of the abudance and clustering of virialized objects 
\cite{virialized}.  In the case that $f_{NL}$ has strong scale dependence cluster
number counts may, in fact, provide better propects for detection than 
CMB observations \cite{scale_dep}.

In \cite{nongaus} a class of inflationary models was constructed based on
nonlocal field theories containing infinitely many derivatives (see also
\cite{lidsey}).  The $p$-adic inflation model \cite{pi} belongs to this class.
In \cite{nongaus} it was shown that nonlocal inflation models can give rise to
observably large nongaussianty and, for $p$-adic inflation, it was estimated
that $f_{NL} \sim 120$ (for a string coupling of order unity). However, the
calculation of \cite{nongaus} only considered the perturbations of the matter
field, rather than also including metric perturbations self-consistently. 
This approximation was necessary in order to circumvent the technical
complications associated with performing cosmological perturbation theory
using Lagrangians with infinitely many derivatives.  This neglect meant that
the calculation of \cite{nongaus} only provided a rough estimate of the size
of $f_{NL}$ and gave no information about the shape or running. Furthermore,
the presence of an infra-red (IR) divergence in the momentum integrals lead to
concerns that the curvature perturbation might not be freezing in on large scales,
as would be expected in models with only adiabatic perturbations \cite{conserved}.

In this paper we reconsider the calculation of the bispectrum in nonlocal
inflationary models, making use of recent advances both in cosmological
pertubation theory \cite{seery} and also in the formal mathematics of infinite
order partial differential equations \cite{niky}.  The analysis is rendered
tractable by using a particular choice of field basis and by taking advantage
of the hierarchy $\epsilon \ll |\eta|$ among the slow-roll parameters in
hill-top inflation models.  The calculation is further simplified by use of
the recent formalism for computing bispectra directly from the field equations
developed by Seery, Malik and Lyth \cite{seery}.  Our analysis confirms the
qualitative estimate of \cite{nongaus} while improving significantly on the
accuracy.  

The organization of this paper is as follows.  In section \ref{Lsec} we
discuss the class of nonlocal theories under discussion and demonstrate how
the Lagrangian may be simplified by use of appropriate field redefinitions. 
In section \ref{pertsec} we write out the perturbed Klein-Gordon equation at
first and second order in  perturbation theory.  In section \ref{bisec} we use
the results of \cite{seery} to construct the nonlinearity parameter for
nonlocal inflation.  In section \ref{psec} we specialize to the case of
$p$-adic inflation, confirming previous claims that $f_{NL}$ can be made large
in that model.  Finally, in section \ref{concl} we conclude.

\section{Nonlocal Hill-Top Inflation}
\label{Lsec}

\subsection{Lagrangian and Field redefinitions}

We are interested in a class of nonlocal scalar field theories of the form
\begin{equation}
\label{L}
  \L = \gamma^4\left[ \frac{1}{2}\phi \Fbox \phi - U(\phi) \right]
\end{equation}
where, following \cite{nongaus}, we take $\gamma$, $m_s$ to have dimensions of energy while $\phi$, $U(\phi)$
and $F(z)$ are dimensionless.  Lagrangians of the form (\ref{L}) are ubiquitous in string field theory
\cite{SFT} and other stringy constructions \cite{padic}-\cite{discrete} and can also be motivated
in certain braneworld models \cite{claudia} or as a low energy description of unparticles 
\cite{unparticles}.  Phenomenological constructions of the form (\ref{L}) have also attracted interest due to possible
implications for the cosmological constant problem \cite{degrav}. (Similar nonlocal constructions are expected to arise generically in 
quantum field theories with a minimal length scale \cite{minimal}.)  In (\ref{L}) the mass scale $m_s$
gives the scale of the underlying nonlocality of the theory.  For models derived from string theory
this will coincide with the string scale $\sqrt{\alpha'}$, however, we do not require this assumption in our analysis.

We assume a hill-top \cite{hilltop} form for the potential
\begin{equation}
\label{pot}
  U(\phi) = U_0 - \frac{\mu^2}{2!}\phi^2 + \frac{g}{3!}\phi^3 + \cdots
\end{equation}
so that $\phi=0$ is an unstable maximum.  In (\ref{pot}) $\cdots$ denotes quartic and higher order 
terms which stabilize the potential at large values of $\phi$.  Since we are primarily interested in
dynamics near the unstable maximum we can neglect the terms $\O(\phi^4)$ and higher.

In order for the theory to be free of ghosts and Ostrogradski instabilities we require that $F(z)$ can be cast in the form
\begin{equation}
\label{F}
  \gamma^2 \left[ \Fbox + \mu^2 \right] = \Gbox ( {\Box} + \omega^2 )
\end{equation}
where $\Gamma(z)$ is an analytic function of the complex variable $z$ having no zeros at finite $z$ \cite{niky}.  
Notice that,
as in \cite{nongaus}, we have 
\be
\mu^2 = -F(-\omega^2 / m_s^2) 
\label{mueq}
\ee
and $F(0) = 0$.  We will be particularly interested in the following special values
of the function $\Gamma(z)$
\begin{eqnarray}
  \Gomega &=& \frac{\gamma^2}{m_s^2} \Fpws \label{Gamma} \\
  \Gamma(0) &=& \frac{\gamma^2 \mu^2}{\omega^2} \label{Gamma_zero}
\end{eqnarray}

The structure (\ref{F}) for the kinetic function $\Fbox$ is required in order to ensure that the propagator has only a single pole and the
resulting quantum theory (\ref{L}) describes only a single degree of freedom.  This constraint also ensures that the Cauchy problem associated
with the classical equation of motion is well-posed with only two initial data \cite{niky}.  (See also \cite{localization} for an alternative approach
to counting initial data.)  Note also that theories of the form (\ref{L}) which
\emph{do} suffer from the presence of ghosts (or classical higher derivative instabilities) may nevertheless be rendered physically sensible by 
appropriately redefining the action pseudo-differential operator $\Fbox$ using the prescription advocated in \cite{niky}.

The equation of motion derived from (\ref{L}) is
\begin{eqnarray}
  \Fbox \phi &=& U'(\phi) \nonumber \\
   \Gbox (\Box + \omega^2) \phi &=& {\gamma^2} \Uint'(\phi) \label{EOM1}
\end{eqnarray}
where $\Uint$ denotes the terms in $U(\phi)$ which are of higher order than quadratic
\begin{equation}
\label{Uint}
  \Uint = \frac{g}{3!}\phi^3 + \cdots
\end{equation}

In order to study perturbations it will be convenient to introduce a ``dressed'' field $\varphi$ (see \cite{dressed}) defined through
a nonlocal field redefinition
\begin{equation}
\label{varphi}
  \varphi  = \gamma \, \Gbox^{1/2} \phi
\end{equation}
We will also refer to $\varphi$ as the canonical inflaton for reasons which will become obvious shortly.
Such a nonlocal field redefinition is allowed since $\Gamma(z)$ contains neither zeroes nor poles \cite{tirtho}, hence there
is no danger of altering the structure of the initial value problem \cite{niky}.  In terms of the dressed field the Lagrangian (\ref{L})
takes the form
\begin{equation}
\label{L2}
  \L = \frac{1}{2}\varphi \left(\Box + \omega^2 \right)\varphi - V_0 - \Vint\left[\Gbox^{-1/2}\varphi\right]
\end{equation}
where $V_0 = \gamma^4 U_0$ and 
\begin{equation}
\label{Vint}
  \Vint\left[x\right] = \gamma^4 \Uint\left[\frac{x}{\gamma}\right]
\end{equation}
When $\Box \phi = -\omega^2 \phi$ (as occurs in the linear theory) then (\ref{varphi}) reproduces the definition of a ``canonical''
field employed in \cite{lidsey} although it differs from the definition used in \cite{nongaus}.
The equation of motion derived from (\ref{L2}) takes the form
\begin{equation}
\label{EOM2}
  (\Box + \omega^2)\varphi = \Gbox^{-1/2}{\Vint'}\left[\Gbox^{-1/2}\varphi \right]
\end{equation}

\subsection{Background Evolution}

In \cite{nongaus} the homogeneous solutions of the theory (\ref{L}) were constructed using an expansion in powers of $e^{\lambda t}$.  In \cite{pi}
the backround solutions of $p$-adic inflation were constructed using a similar expansion and were verified using an independent formalism (the friction-dominated
approximation) which did not rely on small values of $\phi$.  The result is that the background canonical field $\varphi_0(t)$ and the Hubble scale $H(t)$ evolve as
\begin{eqnarray}
  \varphi_0(t) &\cong& A e^{|\eta| H_0 t} \label{phi_bkg}\\
  H(t) &\cong& H_0 - H_2 e^{2|\eta| H_0 t} \label{H}
\end{eqnarray}
where  
\begin{eqnarray}
  3H_0^2 &=& \frac{V_0}{M_p^2} = \frac{\gamma^4 U_0}{M_p^2} \\
  A^2 &=& \gamma^2 \Gamma(-\omega^2 / m_s^2) = \frac{\gamma^4}{m_s^2} \Fpws
\end{eqnarray}
(using eq.\ (\ref{varphi}) and eq.\ (\ref{Gamma})) and
in (\ref{phi_bkg}-\ref{H}) we have defined the slow roll parameter
\begin{equation}
  \eta = -\frac{\omega^2}{3H_0^2}
\end{equation}
(The explicit expression for $H_2$ is given in \cite{nongaus},
{ $H_2 = \omega^2
H_0\Fpws/[4 U_0 m_s^2]$}.)
  Inflation takes place at large negative values of
$t$ so that $e^{|\eta|H_0 t} \ll 1$ throughout.  {The zero of time is chosen
abitrarily in the solution (\ref{phi_bkg}), but this of course has no effect on
any physical observable.}
We define a second slow-roll parameter
\begin{equation}
\label{epsilon}
  \epsilon \equiv -\frac{\dot{H}}{H^2} \cong \frac{\dot{\varphi}_0^2}{2 M_p^2 H^2} 
\end{equation}
The remarkable feature of the solutions (\ref{phi_bkg}, \ref{H}) is that slow-roll inflation proceeds even when the naive potential (which would be inferred by setting $\Box = 0$
in (\ref{L})) is extremely steep.

Notice that, as is typical of hill-top models, we have large hierarchy between the slow roll parameters $\epsilon \ll |\eta|$.  It is straightforward to verify that
\[
  \frac{\epsilon}{|\eta|} \cong 2\frac{H_2}{H_0} e^{2|\eta|H_0 t}  \ll 1
\]
so that in the limit $\epsilon / |\eta| \rightarrow 0$ one has $ e^{2|\eta|H_0 t} \rightarrow 0$ and thus $\varphi_0(t) \rightarrow 0$, $H(t) \rightarrow H_0$.
Following \cite{nongaus} we conservatively take the end of inflation to be when $\epsilon \sim |\eta|$, rather than $\epsilon \sim 1$, since this is the point where
the perturbative expasion (\ref{H}) breaks down.\footnote{The $\O(e^{2|\eta|H_0t})$ correction to (\ref{phi_bkg}) was computed in \cite{nongaus} where it was shown
that this term is typically subdominant to the leading term even when $\epsilon > |\eta|$.}  
(In the case of $p$-adic inflation the friction dominated approach gives the same result for the end of inflation \cite{pi}.)
It follows that it is consistent to work with a pure de Sitter background $\varphi_0(t) = 0$, $H(t) = H_0$ while keeping
terms up to leading order in the $\eta$ slow-roll parameter to study perturbations.

\section{The Perturbed Field Equations}
\label{pertsec}

\subsection{Linear Perturbations}

The computation of the bispectrum for the theory (\ref{L}) is greatly simplified by using the formalism of Seery et al.\ \cite{seery} for working directly with the 
perturbed Klein-Gordon equation.  Since (\ref{L2}) has canonical structure in the limit $\Vint \rightarrow 0$ it is clear that (\ref{EOM2}) is the appropriate starting 
point to match onto the analysis of \cite{seery}.  We perturb the field as
\begin{equation}
  \varphi(t,{\bf x}) = \varphi_0(t) + \delta \varphi(t,{\bf x})
\end{equation}
and work in uniform curvature gauge (neglecting tensor modes) so that the ADM \cite{ADM} metric can be written as 
\begin{equation}
  ds^2 = -N dt^2 + \delta_{ij}\left(dx^i + N^i dt \right)\left(dx^j + N^j dt \right)
\end{equation}
It will be convenient to work in terms of conformal time $\tau$ defined by $a d\tau = dt$ so that, in the limit $\epsilon \rightarrow 0$,
one has $a = -1/(H_0\tau)$.  We denote derivatives with respect to $\tau$ by $f' = \partial_\tau f$ and introduce a conformal time Hubble scale
$\sH = a' / a = \dot{a}$.

Following \cite{seery} we decompose the field perturbation as
\begin{equation}
  \delta \varphi(t,{\bf x}) = \done\varphi(t,{\bf x}) + \frac{1}{2}\dtwo\varphi(t,{\bf x})
\end{equation}
where $\done\varphi$ is defined so that it obeys exactly gaussian statistics.  Since the nonlocal structure of $\Vint$ 
in the Lagrangian (\ref{L2}) does not appear at the linear level, the equation for $\done\varphi$ is precisely the same result
that one would have in a standard (local) model\footnote{In \cite{niky} the equivalence of (\ref{L}) to a local theory at the linearized
level was demonstrated using the formal calculus of pseudo-differential operators.}, derived in \cite{malik2}
\begin{equation}
\label{phi1}
\done\varphi'' + 2\sH \done\varphi' - \lap\done\varphi - a^2\omega^2\done\varphi  - 2\epsilon \frac{a^2 V}{M_p^2}\done\varphi = 0
\end{equation}
The final term in (\ref{phi1}) arises from the terms in $\Box\varphi$ involving the metric perturbations which, to leading order in perturbation theory,
always involve time derivatives of $\varphi_0$ and hence are proportional to $\epsilon$.  In the limit $\epsilon \rightarrow 0$ these terms vanish
and (\ref{phi1}) is identical to the dynamical equation for a light field in de Sitter space.  Denoting the d'Alembertian of de Sitter space by $\Box_0$,
that is
\begin{equation}
\label{Box0}
  \Box_0 = -\partial_t^2 - 3H_0\partial_t + e^{-2H_0 t}\lap
\end{equation}
we see that (\ref{phi1}) is equivalent to 
\begin{equation}
\label{dS}
\Box_0\done\varphi \cong -\omega^2\done\varphi
\end{equation}
for $\epsilon \rightarrow 0$.  In other words, in the limit $\epsilon \ll |\eta|$,
we can safely replace $\Box \done\varphi$ with $\Box_0\done\varphi$ throughout the calculation.  This trivial observation will greatly simplify
our treatment of the nonlocal structure of the theory (\ref{L2}).

The solutions of (\ref{phi1}) are well-known
\begin{eqnarray}
  \done\varphi(t,{\bf x}) &=& \int\frac{d^3k}{(2\pi)^{3/2}}e^{i{\bf k}\cdot {\bf x}}\xi_{\bf k}(t) \label{mode_expansion}\\
  \xi_{\bf k}(t) &=& a_{\bf k} \varphi_{\bf k}(t) + a_{-\bf k}^{\dagger}{\varphi^*}_{-\bf k}(t) \label{quantum_mode}\\
  \varphi_{\bf k}(t) &=& \frac{e^{i\delta}}{2}\sqrt{\frac{\pi}{a^3 H_0}} H_{\nu}^{(1)}\left(\frac{k}{aH_0}\right) \label{classical_mode}
\end{eqnarray}
where the order of the Hankel function is 
\[
  \nu = \sqrt{\frac{9}{4} + \frac{\omega^2}{H_0^2}} \cong \frac{3}{2} + \O(\eta)
\]
and we have normalized the annihilation/creation operators as $\left[a_{\bf k},a_{\bf k'}^{\dagger}\right] = \delta^{(3)}({\bf k} - {\bf k'})$.  
In (\ref{classical_mode}) the constant real-valued phase $\delta$
is irrelevant for our calculation.  To leading order in the $\eta$ parameter we have
\begin{equation}
\label{mode}
  \varphi_k(\tau) \cong \frac{H_0}{\sqrt{2{k^3}}}e^{-ik\tau}(1+ik\tau)
\end{equation}

Using the solution (\ref{mode_expansion}-\ref{classical_mode}) it is straightforward to construct the curvature perturbation in the linear theory.
The analysis is essentially identical to \cite{nongaus}, the only difference being that (\ref{varphi}) implies a slightly different normalization for
the canonical scalar which, in turn, gives a slightly different expression for the COBE normalization.  In the appendix we redo the linear analysis of
\cite{nongaus} taking into account this change of normalization.

\subsection{Second Order Perturbations}

We now turn to the equation for $\dtwo \varphi$.  All of the contributions coming from the left-hand-side of (\ref{EOM2})
appear already in the standard (local) theory and have been computed in \cite{malik1,malik2}.  The only new contribution coming
from the nonlocal structure of (\ref{L2}) arises due to the term $\Vint\left[\Gamma^{-1/2}(\Box / m_s^2)\varphi\right]$ and appears on the right-hand-side
of (\ref{EOM2}).  Thus, at leading order in slow roll parameters, the equation for $\dtwo\varphi$ is
\begin{equation}
\label{phi2}
  \dtwo\varphi'' + 2\sH\dtwo\varphi' - \lap\dtwo\varphi -a^2 \omega^2\dtwo\varphi = \Jinttwo + F_2(\done\varphi) + G_2(\done\varphi)
\end{equation}
where the source terms $F_2$, $G_2$ are the same as those appearing in \cite{seery}.  The quantities $F_2$, $G_2$ were explicitly computed in terms of field perturbations in 
\cite{malik1,malik2} and their contributions to $f_{NL}$ were derived in \cite{seery}.  The only new term in (\ref{phi2}) is $\Jinttwo$ which denotes
the leading contribution in a perturbative expansion of
\begin{equation}
\label{Jint}
  \Jint = -2a^2 \Gbox^{-1/2} \Vint'\left[\Gbox^{-1/2}\varphi\right]
\end{equation}

To evaluate $\Jinttwo$ we first consider the argument of $\Vint'$ in (\ref{Jint}).  As we discussed above, in the limit $\epsilon \ll |\eta|$ we can take $\varphi_0 = 0$ and safely replace
$\Box \done\varphi$ by $\Box_0\done\varphi$ (where $\Box_0$ is the d'Alembertian in de Sitter space, given by (\ref{Box0})).  Thus we have
\begin{eqnarray}
  \Gbox^{-1/2} \varphi &\cong& \Gboxzero^{-1/2} \done\varphi \nonumber \\
  &\cong& \Gomega^{-1/2} \done\varphi \label{approx}
\end{eqnarray}
where the corrections are either proportional to $\epsilon$ or are higher order
in perturbation theory.  On the second line of (\ref{approx}) we have used the fact
that $\Box_0 \done\varphi \cong -\omega^2\done\varphi$ in the small-$\epsilon$ limit (see eq.\ \ref{dS}).

Using (\ref{approx}) and focusing on the cubic term $\Vint(x) \cong g \gamma x^3 / 6$ we find that
\begin{equation}
\label{step}
  \Jint \cong - a^2 \frac{g \gamma}{\Gamma\left(-\frac{\omega^2}{m_s^2}\right)} \Gamma\left(\frac{\Box_0}{m_s^2}\right)^{-1/2} \left(\done\varphi\right)^2
\end{equation}
The challenge is now to compute the action of the nonlocal operator $\Gboxzero^{-1/2}$ on the product $(\done\varphi)^2$.  This is, 
in general, extremely complicated since $(\done\varphi)^2$ is not an eigenfunction of $\Box_0$.  
However, since the perturbation $\done\varphi$ behaves as a light field in de Sitter space, the result will be fairly simple.  To see this,
we explicitly compute the expectation value $\langle\Gboxzero^{-1/2}(\done\varphi)^2 \rangle$.

For any analytic pseudo-differential operator $g(\Box_0)$ we have
\begin{eqnarray}
\langle g(\Box_0)(\done\varphi)^2 \rangle &=& \int \frac{d^3k\, d^3k'}{(2\pi)^3}e^{i\, {\bf k}\cdot {\bf x} }\, g(-\partial_t^2 -3H_0\partial_t - k^2/a^2)\,\langle \xi_{k'}(t)\xi_{k'-k}^\dagger(t) \rangle \nonumber \\
 &=& \int \frac{d^3k\,d^3k'}{(2\pi)^3}e^{i\,  {\bf k}\cdot {\bf x}}\, g(-\partial_t^2 -3H_0\partial_t - k^2/a^2)\, \delta^{(3)}(k)\,|\varphi_{k'}(t)|^2 \nonumber \\
 &=& \int \frac{d^3k}{(2\pi)^3} \,\frac{H_0^2}{2k^3}\, g(-\partial_t^2
-3H_0\partial_t)\,\left(1{+}\frac{k^2}{H_0^2}e^{-2H_0t} \right)  \nonumber \\
 &=&   g(0)\, \int \frac{d^3k}{(2\pi)^3}\, \frac{H_0^2}{2k^3}\,\left(1{+}\frac{k^2}{a^2 H_0^2} \frac{g(2H_0^2)}{g(0)} \right) \nonumber \\
\end{eqnarray}
Thus, we have the identity
\begin{equation}
  \langle g(\Box_0)(\done\varphi)^2(t) \rangle = g(0)\,\langle (\done \varphi)^2(t-\Delta t) \rangle \label{identity}
\end{equation}
where the time shift is
\begin{equation}
\label{Deltat}
  \Delta t = \frac{1}{2H_0} \ln\left[\frac{g(2H_0^2)}{g(0)}\right]
\end{equation}
In (\ref{identity}) the overall factor of $g(0)$ arises due to the large-scale constancy of $\done\varphi$ while the time shift $\Delta t$ reflects the fundamental nonlocality of the underlying operation.
Although, in general, such a time shift could have a significant impact on the calculation it turns out to be unimportant here because the variance $\langle(\done\varphi)^2\rangle \sim H_0^2$
is very close to a constant.  Thus, we can safely approximate $g(\Box_0) (\done\varphi)^2  \sim g(0) (\done\varphi)^2$.  This result is intuitively reasonable because $(\done\varphi)^2$ is approximately
constant on large scales while on small scales it is negligible.
Setting $g(\Box_0) = \Gboxzero^{-1/2}$ we have a time shift
\begin{equation}
  \Delta t = -\frac{1}{4H_0} \ln\left[\frac{\Gamma\left(\frac{2H_0^2}{m_s^2}\right)}{\Gamma(0)}\right]
\end{equation}
which amounts to less than a single e-folding in the case of $p$-adic inflation.

In summary, we find that 
\begin{equation}
\label{Jint_final}
  \Jint \cong - a^2 \frac{g \gamma }{\Gomega \Gamma(0)^{1/2}} \, \left(\done\varphi\right)^2
\end{equation}
which takes exactly the form which would arise from a cubic interaction of 
the form $V_{\mathrm{int}} = {c_V} (\done\varphi)^3$ where the effective cubic coupling is
\begin{equation}
\label{cH}
  {c_V} = {+}\frac{g \gamma}{3!} \frac{1}{\Gomega \Gamma(0)^{1/2}}
\end{equation}
Equations (\ref{Jint_final}) and (\ref{cH}) are the main results of this section.

Notice that (\ref{cH}) does not exactly coincide with the definition of the
quantity ${c_V}$ in \cite{nongaus}, {due to the more quantitative treatment 
we are carrying out here}.  However, for the interesting case of
$p$-adic inflation both results are
the same order of magnitude.

\section{The Bispectrum and Nongaussianity}
\label{bisec}

\subsection{Results for $f_{NL}$}

The result (\ref{Jint_final}) is extremely simple.  Since a term of this form was already included in \cite{seery} it follows that we can obtain $f_{NL}$ simply by substituting $V'=0$,
$V'' = -\omega^2$ and $V''' = 6 {c_V}$ into the results of \cite{seery}.  As in \cite{nongaus}, large nongaussianity will result due to the fact that the effective cubic interaction can be made
extremely large without spoiling slow roll.  
This is a result of the nonlocal structure of the theory; the higher derivative structure in  $\Vint$ can mimic a strongly interacting theory even when the actual couplings are all perturbative.  
(From the perspective of the undressed Lagrangian (\ref{L}) one would say that the nonlocal corrections to the kinetic term conspire to slow the rolling of the inflaton even when $U(\phi)$ is 
extremely steep.)  Notice that the size of $c_V$ is, however, bounded from above by the requirement of perturbative stability against quantum corrections.  This bound is 
discussed in \cite{nongaus}.\footnote{For the case of $p$-adic inflation the bounds of
\cite{nongaus} are equivalent to $g_s \lsim 1$ which is precisely the result
one obtains in the full string theory set-up.}  

In order to make contact with the analysis of \cite{seery} we must construct
and effective value for the slow-roll parameter ${\xi^{(2)}} = M_p^4 V'
V'''/V^2$ which parameterizes the size of $V'''$ in a local inflationary
model.   In our case, however, {$\xi^{(2)}$ can greatly exceed unity without
violating the conditions for slow roll; we will discuss this issue in
section \ref{consistency} .} 

{To compute an effective value of ${\xieff}$ in the nonlocal model, which
will play the same role relative to $f_{NL}$ as does ${\xi^{(2)}}$ in local
models, we need to replace $V'''$ by an effective value  $V'''_{\rm eff}$,
which can be deduced by comparing our result for  $J_{\rm int}$, eq.\
(\ref{Jint_final}), with ref.\ \cite{seery}'s corresponding expression,
$J_{\rm int} = - a^2 V''' \delta_1\phi^2$.  This gives $V'''_{\mathrm{eff}}=6
c_V$.  To compute $V'$ in conventions of ref.\ \cite{seery}, we should work in
terms of the canonically normalized field $\varphi$, so that $V' =
-\omega^2\varphi_0 =  -\omega^2\dot\varphi_0 /(|\eta|H_0) = 
-\dot\varphi_0 V_0/(M_p^2 H_0)$.}  In this way we find that
\begin{equation}
\label{xi}
  \xieff \cong -{{\rm sign}(\varphi_0)}\sqrt{2\epsilon} M_p^3\frac{V'''_{\mathrm{eff}}}{V_0}
\end{equation}
where the slow roll parameter $\epsilon$, defined in (\ref{epsilon}),
is understood to be evaluated at horizon crossing.   
Equation (\ref{xi}) can be rewritten in a form similar also to that given in ref.\ \cite{nongaus}
\begin{equation}
\label{xi2}
  \xieff  \cong \frac{2}{H_0^2 c_\zeta^2} {c_V} c_\zeta \cong \frac{1}{2\pi^2} \frac{1}{A_\zeta^2} c_\zeta {c_V}
\end{equation}
where $c_\zeta$ is defined by (\ref{c_zeta}) and we have used (\ref{COBE1}).  (Equation (\ref{xi2}) should be compared
to eqn.\ 4.15 in \cite{nongaus}.)

In order to give a sense of how the quantity $\xieff$ scales with the various couplings in the theory we substitute the explicit expression for $c_V$ (\ref{cH}) 
into (\ref{xi2}) and use (\ref{Gamma}-\ref{Gamma_zero}) as well as the results of the appendix to rewrite $\xieff$ in the form 
\begin{equation}
\label{par_ex}
  \xieff  \cong -2g\,{\rm sign}(\varphi_0)\frac{M_p^4\omega^2
m_s^2 \exp(N_e M_p^2 \omega^2/ \gamma^4 U_0)}{\mu\gamma^8 U_0^{3/2}
 F'(-\omega^2 / m_s^2)}
\end{equation}
This expression shows explicitly how $\xieff$ depends on the model parameters $\mu$, $U_0$, $g$, $m_s$ $F(z)$ and the derived ``effective mass'' $\omega$.  
However, eq.\ (\ref{par_ex}) does not take into account the various constraints among these quanties which are required for consistency with CMB observations.

To more explicitly evaluate $\xieff$, we would like to incorporate the
experimental contraints on the amplitude and index of the CMB power spectrum.
These constraints allow us to fix two combinations of the model parameters,
which can be chosen as $\mu^2$ and $\gamma^4 U_0$.  It is convenient to express
these in terms of the combination
\be
	{f_A \ =\  12\pi^2 A_\zeta^2 e^{-N_e|n_s-1|} |n_s-1|^2\ \sim\  
(0.1-1)\times 10^{-10}}
\label{fa}
\ee
where $A_\zeta^2 = 25\times 10^{-10}$ is the COBE normalization for the amplitude of the
scalar power spectrum (see the appendix for more details).  The quantity $f_A$ in (\ref{fa})
depends upon observables rather than parameters of the model.  {In figure \ref{fafig} we 
show the dependence of $f_A$ on $n_s$ for a range of $N_e$.}
 Using eqs.\ (\ref{mueq}), (\ref{gamma_identity}) and (\ref{COBE1}), we find
\bea
\label{cons1}
	\mu^2 &=& -F\left(-f_A{M_p^2\over m_s^2}\right)
\ \Longleftrightarrow\ \omega^2 = M_p^2 f_A \\
\label{cons2}
	\gamma^4 U_0 &=& 2 M_p^4 f_A |n_s-1|^{-1}
\eea  
Using these equations to eliminate $\mu^2$ and $U_0$, we find that
\begin{equation}
\label{cHczeta}
  {\xieff  = -{\rm sign}(\varphi_0){g \gamma\over f_A M_p 
	\Gamma(-f_A M_p^2 / m_s^2)}
	\left( { |n_s-1|^3}\over 2\Gamma(0)e^{{N_e}|n_s-1|}\right)^{1/2}}
\end{equation}



\EPSFIGURE{fa.eps,width=4in}{$f_{A}$ versus $n_s$, for $N_e = 40,\dots, 60$.\label{fafig}}


The expression for the nonlinearity parameter can be written simply in terms of $\xieff$ as \cite{seery}
\begin{equation}
\label{fNL}
  f_{NL} = \frac{5}{6}\xieff\left[ N_e +  0.91  + \frac{3}{\sum_i k_i^3}\left( k_t\sum_{i<j}k_ik_j - \frac{4}{9}k_t^3 \right) \right] + \cdots
\end{equation}
{where $k_t = \sum_i k_i$} and $\cdots$ denotes the (subleading) terms which appear in the standard
calculation.  As was explained in \cite{seery}, the dependence of the 
number of e-foldings since horizon crossing $N_e$ does not imply super-horizon evolution of $\zeta$ since this explicit dependence is cancelled
by implicit $N_e$ dependence in the parameters $\xieff$ and $\eta$.\footnote{One may verify using the results of \cite{flow} that the relevant flow equations
continue to hold even when $\xieff$ is larger than unity.  It follows that the same arguments given in \cite{seery} imply that $\zeta$ is frozen on super-horizon scales
in nonlocal inflation, at least to the accuracy of our calculation.}

To estimate the size of the nongaussianity we evaluate (\ref{fNL}) on equilateral triangles $k_1 = k_2 = k_3$
\begin{equation}
  {f_{NL}^{{\triangle}} =  
 \frac56(N_e - 2.09)\xieff}
\label{finalfnl}
\end{equation}
{Equations (\ref{cHczeta}-\ref{finalfnl}) are the main result of this section.  
Using these expressions one may straightforwardly evaluate $f_{NL}$ in terms 
of the model parameters $g$, $U_0$, $\mu$, $m_s$, subject to the constraints
(\ref{cons1}-\ref{cons2}).  Note that $\omega$ is a derived parameter, 
$\omega^2 = -m_s^2 F^{-1}(-\mu^2)$.}  We will confirm that, unlike in standard
single-field slow roll inflation, $f_{NL}$ can be quite large in nonlocal 
models.

Finally, let us comment on our choice of sign convention for $f_{NL}$.  Our definition follows \cite{seery}
and coincides with the convention used by the WMAP team \cite{WMAP5} and also by \cite{detection}, however, it
differs from some other studies \cite{Maldacena,large_nongauss}.  Notice that the final result for $f_{NL}$
in \cite{seery} actually has a sign error for the terms involving $\epsilon_\star$ and $\eta_\star$, however, this error does
not affect the $\xi^2_\star$ term coming from the cubic interaction.\footnote{We are grateful to D.\ Seery for clarification on this point.}  
Since it is this latter term that dominates our result (\ref{fNL}) it follows that our results can be directly compared 
to \cite{WMAP5,detection}.  For a careful discussion of sign
conventions for $f_{NL}$ the reader is referred to appendix A.2 of
\cite{scale_dep}.

\subsection{Consistency of the Slow-Roll Dynamics} 
\label{consistency}

Even in a conventional local model, one could always arrange for $V'''$ to be
anomalously large at some special field value $\varphi_c$, and seemingly obtain a
sizeable contribution to the nongaussianity.  However this would not be
consistent, because the conditions for slow roll would be spoiled for
$\varphi\neq\varphi_c$, leading to a rapid end of inflation, long before the
required number of e-foldings.  We need to verify that the same problem does
not occur in the nonlocal models under consideration.  This is a somewhat
subtle issue, and one can get the wrong answer by being too naive.  

The crucial question which we must address is whether the large values of $\xieff$
required to get large $f_{NL}$ will invalidate the slowly-rolling backgound solutions
(\ref{phi_bkg}-\ref{H}).  In the case of $p$-adic inflation we have verified both the 
background solutions (\ref{phi_bkg}-\ref{H}) and also the expression for the end of inflation 
$\epsilon \sim |\eta|$ using the friction-dominated approximation \cite{pi}.  This formalism 
is nonperturbative and gives a powerful consistency check on the slow roll dynamics of $p$-adic 
inflation in the $|f_{NL}| \gg 1$ regime.  

However, for a general model of the form (\ref{L}) it may be nontrivial to verify the consistency
of the small $e^{|\eta|H_0 t}$ expansion in (\ref{phi_bkg}-\ref{H}) when $|f_{NL}| \gg 1$.
This is so because $|f_{NL}| \gg 1$ tends to require $|g| \gg 1$ and
it is conceivable that terms 
involving higher powers of $e^{|\eta|H_0 t}$, which were neglected in (\ref{phi_bkg}-\ref{H}), 
become important before $\epsilon \sim |\eta|$.  We caution the reader
that, although we present our
results in the most general context, the consistency of the slow roll dynamics should be verified
for any particular model under consideration, as we have done for $p$-adic inflation.

\section{$p$-adic Inflation}
\label{psec}

The Lagrangian of $p$-adic string theory is
\cite{padic}\footnote{{Note that ref.\ \cite{nongaus} missed the factor of
$2$ in the exponent $p^{-\Box/2 m_s^2}$}.  This typographical error was not propagated into the final results.}
\begin{equation}
\label{pL}
  \L = \frac{m_s^4}{g_p^2}\left[-\frac{1}{2}\psi p^{-\Box / {2m_s^2}}\psi + \frac{1}{p+1}\psi^{p+1}\right]
\end{equation}
with $g_p^2 \equiv g_s^2(p-1) / p^2$ and one must take $\psi = 1 + \phi$ to put this in the same form as (\ref{L}).  
The constants $g_s$, $m_s$ are the string coupling and string mass respectively and $p$ is a prime number that characterizes the 
field of numbers in which the worldsheet coordinates take values. (A
$p$-adic number $q$ can be expressed in the form of a series
$q=\sum_{n=k}^\infty c_n
p^n$ where $k$ is a possibly negative integer, $k>-\infty$.)
In \cite{nongaus} expressions for the parameters $U_0$, $g$, $\omega$, $\mu$, 
$\cdots$ 
in terms of $p$, $g_s$ and $m_s$ were derived.  
They are given by
\be
	\mu^2 = p-1,\quad \omega^2 =  {2m_s^2},\quad \gamma^4 = {p^2 m_s^4\over
	(p-1) g_s^2},\quad U_0 = {p-1\over 2(p+1)},\quad g = -\frac12 p(p-1)
\ee
and the kinetic function is $F(z) = 1-p^{-z/2}$.  Using the results (\ref{Gamma},\ref{Gamma_zero})
we obtain the relevant values
\be
	\Gamma(-\omega^2/m_s^2) = {p\ln p\over {2}  g_p} 
	= {p^2\ln p\over  {2} g_s \sqrt{p-1}}
,\quad \Gamma(0) = {p\sqrt{p-1}\over  {2} g_s}
\label{pparams}
\ee
The experimental constraints (\ref{cons1}-\ref{cons2}) then show that
\bea
	m_s &=& \frac{f_A^{1/2}}{\sqrt{2}} M_p,\nonumber\\
	 {(p+1) g_s^2\over p^2} &=& \frac{1}{{16}} f_A |n_s-1|
\label{pcons}
\eea 
We will use (\ref{pcons}) to eliminate $m_s$ and $g_s$, leaving $p$ as the only
free parameter.  Using the values (\ref{pparams}) in (\ref{cHczeta}) 
we find that (again for equilateral triangles in $k$-space)
\begin{equation}
\label{pfNL}
  f_{NL}^{{\triangle}} = \frac{5(N_e - 2)}{24\sqrt{2}}|n_s-1|^2 
e^{-\frac{N_e}{2}|n_s-1|}\, \frac{1}{\ln p}\frac{p-1}{\sqrt{p+1}}
\end{equation}
which remarkably has no dependence on the small parameter $f_A$, and which
scales as $f_{NL} \sim p^{1/2} / \ln p$ in the interesting region of parameter
space $p \gg 1$ where the nonlocal structure of the theory is  playing an
important role in the dynamics.  (This is slightly suppressed compared to the
result of \cite{nongaus} by the factor $\ln p$ although the order unity
prefactors are slightly larger so that the net result has a very similar
magnitude.)   In the limit $p \rightarrow 1$ the $p$-adic theory (\ref{pL})
reduces to a local field theory and (\ref{pfNL}) gives $|f_{NL}| \sim
|n_s-1|$, as expected.  But in the opposite regime of $p\gg 1$, we can  obtain
large nongaussianity, as illustrated in figure \ref{fnlfig}. We have also
evaluated $f_{NL}$ as a function of $g_s$ for the WMAP5-preferred value of
$n_s = 0.96$, and the range $N_e=40-60$ e-foldings, plotting the  result in
figure \ref{fnlgs}.  The curves are approximately linear, and can be well-fit
in the region of observational interest by 
\begin{equation}
  f_{NL}^{{\triangle}} \cong  (23 + 6.76 N_e) g_s
\end{equation}
It shows that for reasonable values of the string
coupling, ${0.15} < g_s < {0.25}$,  we reproduce the central value for $f_{NL}$
claimed by ref.\ \cite{detection}.  The corresponding values of  $p\sim
10^{12}$ as a function of $g_s$ from eq.\ (\ref{pcons}) are shown in figure
\ref{pfig}.  The large value of $p$ (as long as $g_s$ is not extremely small)
is a consequence of the smallness of $f_A\sim A^2_\zeta$ which enters through the
COBE normalization in (\ref{pcons}).  A similar situation occurs in the
DBI inflation models \cite{DBI}, where the dimensionless 't Hooft coupling
must be taken to be $\lambda \sim 10^{12} g_s$.  






\EPSFIGURE{fnl.eps,width=4in}{$f^{\triangle}_{NL}$ for equilateral triangles, as a function of $n_s$, for 
the same values of $N_e = 40,\dots 60$ as in figure \ref{fafig}, and values of the string coupling $g_s= 0.05,\ 0.15,\ 0.5$.\label{fnlfig}}

\EPSFIGURE{fnlgs2.eps,width=4in}{$f^{\triangle}_{NL}$ versus  as a function of $g_s$, for $n_s=0.96$
and  $N_e = 40,\dots 60$.  Horizontal lines show central value and 95\%
c.l.\ range for detection claimed by ref.\ \cite{detection}.\label{fnlgs}}

\EPSFIGURE{p2.eps,width=4in}{$p$ versus $g_s$ for $n_s=0.96$ and $N_e = 40,\dots 60$.\label{pfig}}

It is worth noting that the large $f_{NL}$ in $p$-adic inflation is  {not}
coming solely from the factor $N_e$ in (\ref{fNL}).   Indeed, for $g_s \sim 1$
we still have $f_{NL} \sim 8$ - almost four orders of magnitude larger than in a local field theory -
taking only $N_e =3$.
Rather, the large nongaussianity arises purely due to the fact that the potential $U(\phi)$ is
quite steep and the cubic coupling is large when $p$ is large: $|g| \sim p^2 \gg
1$ in  the $p$-adic model.

An interesting question is whether the result (\ref{fNL}) can be
observationally distinguished from the prediction for models such as DBI
inflation.  In ref.\ \cite{large_nongauss} it was shown that the DBI
model predicts a shape proportional to 
\be
	A_c \propto \left(-{1\over k_t S}\sum_{i>j}k_i^2 k_j^2 +
{1\over {2} k^2_t S}
	\sum_{i\neq j}k_i^2 k_j^3+ \frac18\right) 
\ee
where $S =  \sum_i k_i^3$, 
which differs from the shape predicted for the nonlocal models, (\ref{fNL}).
Let us call the latter $A_{NL}$, and normalize $A_c$ and $A_{NL}$ such that
they are equal for equilateral triangles ($k_1=k_2=k_3$).  A measure for how
easily the two shape dependences could be distinguished from each other is the quantity
\be
	R = { A_{NL} - A_c\over  |A_c| + |A_{NL}|}
\label{Req}
\ee
Both $A_c$ and $A_{NL}$ depend only on the shape and not the size of the
triangles defined by $\vec k_1$, $\vec k_2$ and $\vec k_3 = -(\vec k_1 + \vec
k_2)$.  Thus we can take $R$ to be a function of any two angles, call them
$\alpha_{1,2}$. 
$R$ also depends on  $N_e$ via $A_{NL}$.  Setting $k_3=1$, we
can write $k_i = \sin\alpha_i/\sin(\alpha_1+\alpha_2)$. 
 The physical values of $\alpha_i$ are within the triangle
shown in figure \ref{cont}, whose boundaries represent degenerate limits:
if either $\alpha_i$ vanishes, then two of the vertices coincide, while if
$\alpha_1+\alpha_2=\pi$ then all three vertices lie on a line, but do not
in general coincide.  
 We plot contours of
$R$ in figure \ref{cont} for $N_e=55$, but we find that their shape is quite
insensitive to the value of $N_e$, in the range $40$ to $60$,  due to the
fact that $N_e$ dominates over the $k$-dependent terms in eq.\ (\ref{fNL}),
so that $A_{NL}$ is nearly independent of angles. On the other hand, $A_c$
shows strong dependence on the angles, and the fact that $R$ deviates
significantly from zero over a large range of angles
indicates that the search for this shape-dependence could be a strong
discriminator between nongaussianity coming from nonlocal inflation and that coming 
small sound speed models inflation models.




\EPSFIGURE{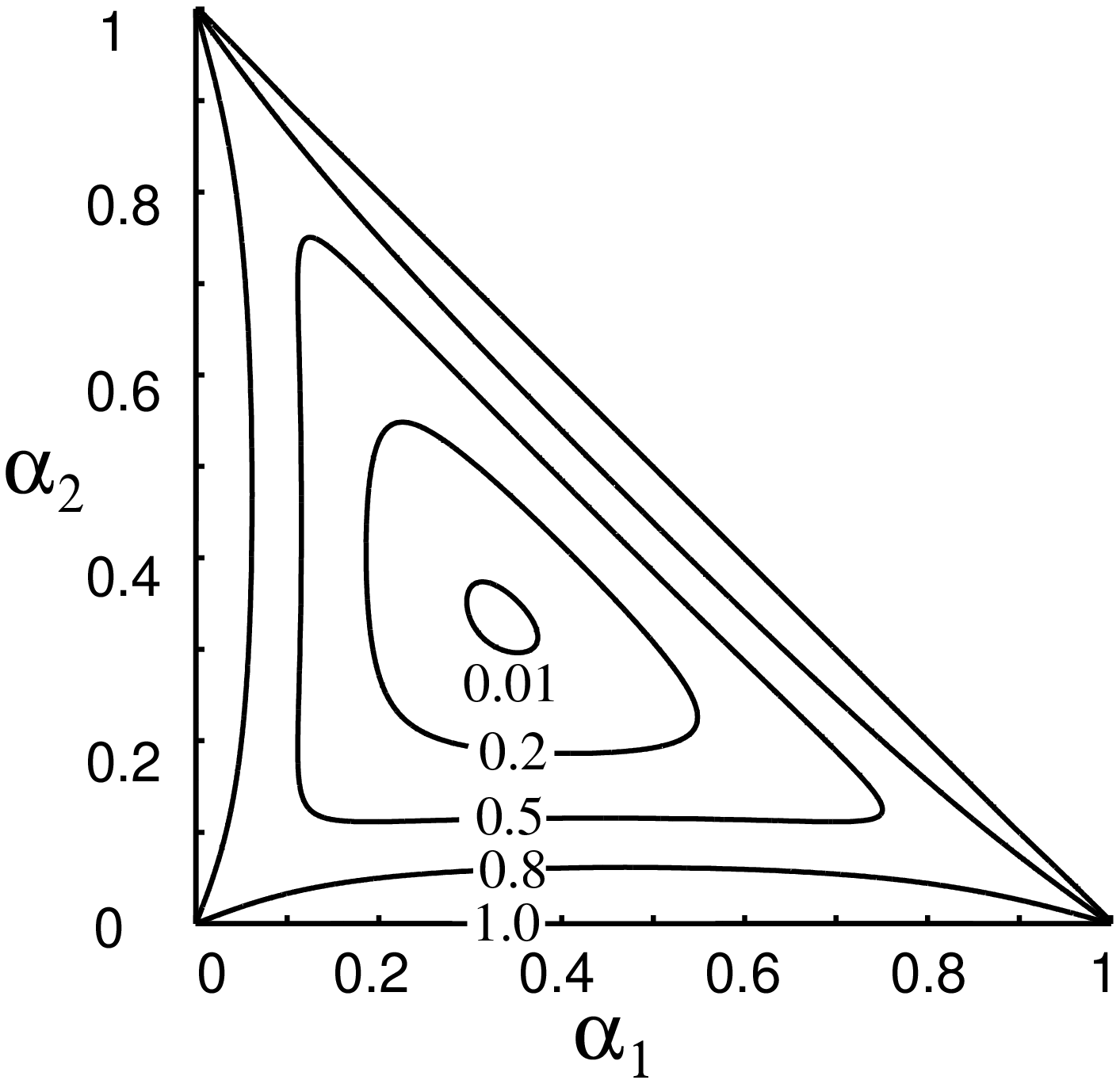,width=4in}{Contours of the ratio $R$, eq.\ 
(\ref{Req}), 
quantifying the
relative shape dependence of nonlocal versus DBI nongaussianity as a function
of the shape of triangles in $k$ space.  The angles are defined by
$k_i/k_3  = \sin\alpha_i/\sin(\alpha_1+\alpha_2)$.\label{cont}}

Using the bispectrum alone it may, however, be more challenging to discriminate observationally between nonlocal inflation and the
curvaton model \cite{curv}, since the latter predicts $f_{NL}$ of the same form as the leading term in (\ref{fNL}) (proportional to $N_e$) \cite{curvaton}.  
A similar analysis to the one described above for DBI inflation shows that the quantity $R$, which disciminates between nonlocal inflation and the curvaton model,
will be at the percent level when $N_e \sim 30-60$ ($R$ could be larger for smaller values of $N_e$).  However, we note that if $|f_{NL}| \gg 1$
is observed then it may be much easier to distinguish between the two models using other constraints, such as those coming from isocurvature
modes \cite{iso} which should be significant in the curvaton model when $|f_{NL}| \gg 1$ but are absent in nonlocal inflation.

Notice that, because we have normalized $A_{NL} = A_c$ on equilateral triangles, it is irrelevant for our comparison that \cite{large_nongauss} uses
a different sign convention for $f_{NL}$ than we do: the quantity $R$ only contains meaningful information about relative shapes of the two bispectra 
but not the amplitude or sign.

\section{Conclusions}
\label{concl}

In this paper we have reconsidered the calculation of the bispectrum in nonlocal hill-top inflationary models.  We have confirmed previous
claims \cite{nongaus} that such constructions can give rise to observably large nongaussianity in the CMB while significantly improving on the 
accuracy of previous estimates.  Large nongaussianity seems to be a generic prediction of nonlocal hill-top inflation in the sense that
$f_{NL}$ tends to be large whenever the nonlocal structure of the theory is playing a significant role in the inflationary dynamics.

We have shown that nonlocal hill-top inflation gives a distinctive prediction for the shape of the nonlinearity parameter which should
be observationally distinguishable from other models which give $|f_{NL}| \gg 1$, such as DBI inflation (or similar models where large nongaussianity
is generated due to a small sound speed for the inflaton).  
It is more difficult to observationally distinguish between nonlocal inflation and the 
curvaton model using only the bispectrum; however, it might be possible using constraints on isocurvature modes.  

In the interesting case of $p$-adic inflation \cite{pi} we have shown that
the predicted value of $f_{NL}$ can be quite close to the central value 
of the claimed detection 
in \cite{detection}, using natural values for the string coupling.
Indeed, future missions should be able to robustly rule out (or confirm!)
the most interesting region of parameter space for this model.  Since
$p$-adic string theory is not construed as a realistic model by itself, it
remains an open problem to construct similar nonlocal inflationary models
in more realistic string theory constructions (addressing also issues of
compactification and moduli stabilization).  On this front we are
optimistic because the nonlocal structure (\ref{L}) is ubiquitous in string
field theory and can be motivated also in a variety of other contexts.

From the formal perspective, one of the most interesting features of
theories of the form (\ref{L}) is that the both the consistency of the
underlying quantum field theory, and also the prediction $|f_{NL}| \gg 1$
in the inflationary context, rely on UV completion.  In other words, if the
Taylor series  expansion of the kinetic function $F(z)$ is truncated at
some finite order $\O(z^N)$ then the resulting truncated theory will, in
general, contain spurious  ghost-like degrees of freedom which are not
present in the full theory (\ref{L}) defined by the exact
transcendental expression for $F(z)$ \cite{niky}.   Moreover, the dynamics
of the truncated theory may be quite different from those of the fully
nonlocal theory and, in general, our prediction for $f_{NL}$  (\ref{fNL})
will not be reproduced.  Remarkably, this statement continues to be true
even for $N \gg 1$.  One might consider the requirement of UV completion 
daunting were it not for the fact that constructions of the form (\ref{L})
can be microscopically justified from string theory.  From this perspective
we  have the exciting possibility that observations of the primordial
bispectrum might be used to probe physics in the quantum gravity regime.

\section*{Acknowledgments}

This work is supported in part by NSERC and FQRNT.  We thank G.\ Smoot 
for discussions about detectability of nongaussianity in future experiments.
We are also grateful to X.\ Chen, J.\ Khoury, D.\ Seery and S.\ Shandera for 
useful discussions and correspondence.

\renewcommand{\theequation}{A-\arabic{equation}}
\setcounter{equation}{0}  

\section*{APPENDIX: Constructing the Linear Curvature Perturbation}

In this appendix we redo the linear computation of the curvature perturbation in \cite{nongaus} taking into account the change of normalization implied by (\ref{varphi}).  Since, for the linear
equation (\ref{dS}) we have $\Box_0 \done\varphi \cong -\omega^2\done\varphi$ equation (\ref{varphi}) simply gives
\begin{equation}
  \done\varphi = A \done\phi
\end{equation}
where, using the identity (\ref{Gamma}), we have
\begin{equation}
\label{A}
  A^2 = \gamma^{2} \Gomega = \frac{\gamma^4}{m_s^2} \Fpws
\end{equation}
from (\ref{varphi}) and (\ref{Gamma}).  The power spectrum associated with the solution (\ref{mode_expansion}-\ref{classical_mode}) is well-known
\begin{equation}
\label{varphi_pwr}
  P_{\varphi} = \left(\frac{H_0}{2\pi}\right)^2\left(\frac{k}{aH_0}\right)^{n_s-1}
\end{equation}
with spectral index
\begin{equation}
\label{ns}
  n_s - 1 = 2\eta = -\frac{2\omega^2}{3H_0^2}
\end{equation}
so that $n_s < 1$, in agreement with the latest WMAP data \cite{WMAP5}.  Using the fact that $3H_0^2 = \gamma^4 U_0 / M_p^2$ equation (\ref{ns})
implies the useful relation
\begin{equation}
\label{gamma_identity}
  \frac{1}{\gamma^4} = \frac{|n_s-1|}{2}\frac{U_0}{M_p^2 \omega^2}
\end{equation}

We now construct the curvature perturbation.  In the comoving gauge we have \cite{malik1}
\begin{equation}
\label{zeta}
  \zeta = -\frac{H}{\dot{\varphi}_0} \done\varphi
\end{equation}
Following \cite{nongaus} we define the constant
\begin{equation}
\label{c_zeta}
  c_\zeta =  -\frac{H}{\dot{\varphi}_0}
\end{equation}
with the understanding that the right-hand-side is evaluated at horizon crossing.  The COBE normalization
fixes the amplitude $A_\zeta$ of the power spectrum of the curvature perturbation
\begin{equation}
\label{zeta_pwr}
  P_\zeta = A_\zeta^2\left(\frac{k}{aH_0}\right)^{n_s-1}
\end{equation}
to be $A_\zeta^2 = 25\times 10^{-10}$.  From (\ref{zeta_pwr}) and (\ref{zeta}) it is clear that we have the relation
\begin{equation}
\label{COBE1}
  c_\zeta^2 H_0^2 \cong (2\pi)^2 A_\zeta^2
\end{equation}

We now evaluate (\ref{c_zeta}).  Using the background solution of \cite{nongaus} we have
\begin{equation}
\label{c_zeta_sqr}
  c_\zeta^2 = \frac{H_0^2}{A^2 \lambda^2}\frac{1}{u_\star^2}
\end{equation}
where $u = e^{\lambda t}$, $\lambda \cong |\eta| H_0$ and $u_\star$ denotes $u(t)$ evaluated at the time
of horizon crossing.  Noting that $u(t) \cong a(t)^{|\eta|}$ and $|\eta| = |n_s-1|/2$ we have 
\begin{equation}
\label{ustar}
  u_\star = e^{-\frac{N_e}{2}|n_s-1|} u_{\mathrm{end}}
\end{equation}
where $N_e$ is the number of e-foldings between horizon crossing and the end of inflation (typically $N_e \sim 60$).
We conservatively take inflation to end when $\epsilon \sim |\eta|$ since this is when the expansion in powers of $u(t)$ becomes invalidated.  This gives (eqn. 2.28 of \cite{nongaus})
\begin{equation}
\label{uend}
  \frac{1}{u_{\mathrm{end}}^2} = \frac{1}{4 U_0}\left(\frac{\omega^2}{m_s^2}\right) F'\left(-\frac{\omega^2}{m_s^2}\right)
\end{equation}
Plugging (\ref{A}), (\ref{ns}), (\ref{ustar}) and (\ref{uend}) into (\ref{c_zeta_sqr}) we derive the useful identiy
\begin{equation}
\label{c_zeta3}
  c_\zeta^2 = \frac{\omega^2}{\gamma^4 U_0}\frac{e^{N_e|n_s-1|}}{|n_s-1|^2}
\end{equation}
Using (\ref{c_zeta3}) and the COBE normalization (\ref{COBE1}) we get an expression for $m_s / M_p$
\begin{equation}
\label{ms}
  \left(\frac{m_s}{M_p}\right)^2 = 12\pi^2A_\zeta^2 |n_s-1|^2 e^{-N_e|n_s-1|} \frac{m_s^2}{\omega^2}
\end{equation}

For the particular case of $p$-adic inflation equations (\ref{ns}), (\ref{COBE1}) and (\ref{c_zeta_sqr}) imply
the following relation between $p$ and $g_s$
\begin{equation}
\label{padic_COBE}
  \frac{4}{3\pi^2}\, \frac{g_s^2(p+1)}{p^2}\, \frac{e^{N_e|n_s-1|}}{|n_s-1|^3} = A_\zeta^2
\end{equation}
which is almost identical to eq.\ (5.20) in \cite{pi}.  It follows that this change in normalization is not quantitatively
significant in the case of $p$-adic inflation.


\bibliographystyle{apsrmp}
\bibliography{rmp-sample}

\end{document}